\documentclass[prc,aps,groupedaddress,superscriptaddress,nofootinbib,showpacs
,preprintnumbers,onecolumn]{revtex4}
\usepackage{graphicx}
\usepackage{amsmath} 
\usepackage{amsfonts}
\usepackage{amssymb}
\usepackage{natbib}
\usepackage{dcolumn}
\usepackage{bm}

\begin{document}

\title{Nuclear matter saturation in a relativistic chiral theory and QCD
susceptibilities}
\author{G. Chanfray}
\email{chanfray@ipnl.in2p3.fr}
\affiliation{IPN-Lyon, IN2P3-CNRS et UCB Lyon I, F-69622 Villeurbanne Cedex}
\author{M. Ericson}
\email{magda.ericson@cern.ch}
\affiliation{IPN-Lyon, IN2P3-CNRS et UCB Lyon I, F-69622 Villeurbanne Cedex}
\affiliation{Department of Physics, Theory Division, CERN, CH-12111 Geneva}
\date{\today}
\begin{abstract}
We study a chiral relativistic theory of nuclear matter aimed at the desciption 
of both the binding and
saturation properties and the QCD properties, quark condensate and QCD susceptibilities. 
For this  purpose the nucleon scalar response of the quark-meson coupling model 
is introduced in the linear sigma model. The consequences for the nuclear and the QCD 
scalar susceptibilities are discussed. 
\end{abstract}
\maketitle

 \section{Introduction}
The order parameter associated with spontaneous breaking of chiral symmetry, namely  
the chiral quark condensate,  is influenced in the nuclear medium by the mean scalar 
field which possesses the same quantum numbers \cite{CEG02}.   
Its  fluctuations in the medium as well are intimately  related to the in-medium 
propagation of the scalar field \cite{CE03}. On the other hand   the properties of 
the scalar field influence in a crucial way the question of the nuclear binding. 
It is therefore important for both problems to reach a description of the nuclear dynamics 
which satisfies the chiral constraints and is able to correctly reproduce the binding and 
saturation properties. The key ingredients can be found in the theory of quantum  
hadrodynamics (QHD) \cite{SW86,SW97}, which,  however, in its original form 
lacks the chiral aspects. We have recently shown that, starting from a chiral effective
model, it is possible to give a chiral status  to QHD  by imposing a chiral invariant 
character to the scalar field of QHD  \cite{CEG02}. 
However there is a well identified problem concerning the nuclear saturation with usual 
chiral effective theories \cite{BT01}. Independently of the particular chiral model, 
in the nuclear medium one moves away from the minimum of the vacuum effective potential 
(mexican hat potential), {\it i.e.}, into a region of smaller curvature. This single effect,
equivalent to the lowering of the sigma mass, destroys the stability, creating problems 
for the applicability of 
such effective theories in the nuclear context. The problem can be cured with the 
introduction of the nucleonic response to the scalar field which is the central ingredient 
of the quark-meson coupling model, QMC \cite{G88,GSRT96}. Indeed, the nucleon 
polarization under the influence of a scalar field implies a density dependence of the 
scalar nucleon coupling constant, which actually decreases with increasing density. This 
effect can counterbalance the decrease of the curvature of the mexican 
hat potential and restore saturation. The QMC model is quite successful in the phenomenology 
of nuclear matter or finite nuclei.  It is the aim of this work to incorporate its  
main concept, 
{\it i.e.}, that of a nucleonic response to a scalar field, in an  effective chiral 
theory, the linear sigma model. This concept is the only feature that we retain of QMC, which 
goes beyond  since it evaluates the scalar polarizability $\kappa_{NS}$ in a bag model. For 
us $\kappa_{NS}$ is  a parameter that we can adjust. We are however guided by the values 
obtained in the original model. In particular the sign found in QMC is crucial. The
fact that the nucleonic scalar response manifests itself at the nuclear level is plausible. 
This is familiar in other situations. For instance under the influence of an isovector axial 
field, the  nucleon converts into a Delta. This axial polarizability of the nucleon has 
several physical implications (such as the in-medium renormalization of the axial 
coupling constant). What is unusual in the scalar polarizability is its sign. In QMC, 
the scalar polarizability originates from relativistic effects of quarks confined in a bag 
(Z graphs) and it is positive. It screens the scalar field (diamagnetic effect).

Inside of our framework, {\it i.e.}, the linear sigma model with a  nucleonic scalar
response, we will choose the parameters so as to be compatible
with the known saturation properties of nuclear matter: saturation density, binding 
energy and compression modulus. The resulting density dependence of the sigma mass becomes
quite mild while, in the absence of the nucleonic scalar response, this mass drops rapidly 
with increasing density,  preventing saturation to occur. We also derive the QCD scalar 
susceptibility of the nuclear medium. An interesting by-product of our work is the obtention 
of the residual  particle-hole interaction in the isoscalar channel. It arises from in-medium
sigma and omega exchanges. The introduction of the nucleonic scalar response 
leads to a rapid density dependence of this interaction. It turns from attraction to 
repulsion near saturation density. Accordingly the collective nature of the response 
undergoes a qualitative change, from a softening effect at low density to a hardening 
one above saturation density.

Section 2 is devoted to the formal derivation from the equation of state of various 
quantities: sigma mass, QCD susceptibility, nuclear response. In section 3 we perform 
the numerical evaluations of these quanties, ensuring the compatibility 
with the nuclear phenomenology  and we discuss the consequences.

 \section{Equation of state and scalar fluctuations}
We start with the Lagrangian introduced in our previous work \cite{CEG02}, dropping  its 
pionic part which does not contribute at the mean field level:
\begin{equation}
{\mathcal{L}}={1\over 2}\partial ^{\mu }s \partial
_{\mu }s \, -\, V(s)\,+\,i\bar{N}\gamma ^{\mu }\partial _{\mu }N\, -
\, M_{N}\left( 1+{s\over f_{\pi }}\right) \bar{N}N\
\end{equation}
where $s$ is the chiral invariant field associated with the 
radius $S=s\,+\,f_\pi$ of the chiral circle. $V(s)=V_0(s)\,-\, c S$ is the vacuum
potential which can be split into $V_0(s)=(\lambda/4)\big((f_\pi+s)^2-v^2\big)^2$ 
responsible for spontaneous chiral symmetry 
breaking and the explicit symmetry breaking piec, $- c S$, where $c= f_\pi m_\pi^2$. 
As usual in QHD we  add a coupling to an omega field~: 
\begin{equation}
{\mathcal{L}}_V={1\over 2}\partial ^{\mu }\omega \partial_{\mu }\omega\,+\,
{1\over 2}\, m^2_\omega\,\omega^2\,-\, g_\omega\,\omega\, N^\dagger N .
\end{equation}
As for the nucleonic  response, $\kappa_{NS}$, to the scalar field $s$, 
we incorporate it in the following extra term in the lagrangian~: 
\begin{equation}
{\mathcal{L}}_\chi= - {1\over 2}\,\kappa_{NS}\, s^2\,\bar{N}N .
\end{equation}
The susceptibility $\kappa_{NS}$ embeds the influence of the 
internal nucleon structure. We will discuss later its possible $s$ dependence. 
At the mean field level the energy density is given by~:
\begin{equation}
\varepsilon=\int\,{4\,d^3 p\over (2\pi)^3} \,\Theta(p_F - p)\,E^*_p(\bar s)
\,+\,V(\bar s)\,+\,{g^2_\omega\over 2\, m_\omega^2}\,\rho^2
\end{equation}
where the baryonic density is related to the Fermi momentum through~:
\begin{equation}
\rho=\int\,{4\,d^3 p\over (2\pi)^3} \,\Theta(p_F - p)
\end{equation}
and $E^*_p(\bar s)=\sqrt{p^2\,+\,M^{*2}_N(\bar s)}$ is the energy of an 
effective nucleon with the effective mass~:
\begin{equation}
M^*_N=M_N\,\left( 1+{\bar s\over f_{\pi }}\right)\,+{1\over 2}\,\kappa_{NS}\,
\bar s^2 .
\end{equation}
The expectation value, $\bar S= f_\pi + \bar s$, of the $S$ field, plays the role 
of the chiral order parameter. It is obtained by minimizing the energy density~:
\begin{equation}
{\partial\varepsilon\over \partial\bar s}= g^*_S\,\rho_S\,+\,V'(\bar s)=0
\label{MIN}\end{equation}
with the following expressions for the scalar density, $\rho_S$, 
and the scalar coupling constant $g^*_S$~:  
$$\rho_S=\int\,{4\,d^3 p\over (2\pi)^3} \,\Theta(p_F - p)\,{M^*_N\over E^*_p}
\quad\hbox{and}\quad g^*_S(\bar s)={\partial M^*_N\over\partial\bar s}={M_N\over f_{\pi}}
\,+\,\kappa_{NS}\, \bar s . $$
Notice that the density dependence of $g^*_S$ entirely arises from the susceptibility
term. Since the mean scalar field is negative and the sign of  $\kappa_{NS}$ positive, 
$g^*_S$ is a decreasing function of the density.  In the vacuum the scalar coupling constant 
of the model is $g_S=M_N/f_\pi$.
The in-medium sigma mass is obtained as the second derivative of the energy density
with respect to the order parameter~:
\begin{eqnarray}
m^{*2}_\sigma &=&{\partial^2 \varepsilon\over\partial\bar s^2}=V''(\bar s)\,+\,
{\partial\left(g^*_S\,\rho_S\right)\over \partial \bar s}\nonumber\\
&=&m^{2}_\sigma\left(1 \,+\,{3\bar s\over f_\pi}\,+\,{3\over 2}\left({\bar s\over
f_\pi}\right)^2\right)\,+\,\kappa_{NS}\,\rho_S\,+\,g^*_S\,
{\partial \rho^*_S\over\partial \bar s}
\label{MSIGMA}
\end{eqnarray}
where in the second line we have taken $V"(\bar s)$ in the chiral limit. 
The mean scalar field $ \bar s$  being negative,
the term linear in $\bar s$ (which appears from the curvature of the 
effective potential) in itself lowers the sigma mass by an appreciable amount 
($\simeq 30$ \%  at $\rho_0$). 
This is the chiral dropping  associated with chiral restoration emphasized
by Hatsuda {\it et al} \cite{HA99}, which they suggested to be the origin of the strong 
medium effects found in  $2\pi$ production experiments \cite{B96,ST00,M02}. 
However the scalar susceptibility term (second term in $\kappa_{NS}$ 
on the r.h.s of eq. \ref{MSIGMA}) which is numerically important, 
counterbalances this in-medium mass dropping. Indeed   in QMC, 
the sign of $\kappa_{NS}$  is positive, an essential element of the model for
reaching saturation. The influence of the nucleonic scalar susceptibility thus increases the 
sigma mass, counterbalancing the chiral restoration effect. In pure quantum-hadrodynamics 
instead the chiral softening of the sigma mass is ignored and saturation is obtained through a
 large
 polarisation of the Dirac sea of nucleons. Hence the sigma mass evolution has 
to be discussed in connection with  saturation properties. As for the last term 
($g^*_S \,\partial\rho_S/\partial \bar s$), it also writes~:
\begin{equation}
g^*_S\,{\partial \rho^*_S\over\partial \bar s}
=g^{*2}_S\,\int\,{4\,d^3 p\over (2\pi)^3} \,\Theta(p_F - p)\,{p^2\over E^{*3}_p}.
\end{equation}
We have shown in a previous work  \cite{CEG03} that it actually corresponds to the nuclear 
response associated with $N \bar N$ excitation. 
In practice it is small and it can be omitted.

We will now derive the in-medium chiral condensate and the QCD scalar susceptibility.
They are  related to the first and second derivatives  of the grand
potential with respect to the quark mass $m$ at constant chemical potential $\mu$. 
The baryonic chemical potential is obtained as~:
\begin{equation}
\mu={\partial\varepsilon\over \partial\rho}= E^*_F\,+\,{g^2_\omega \over m_\omega^2}
\,\rho\qquad\hbox{with}\qquad E^*_F=\sqrt{p_F^2\,+\,M^{*2}_N(\bar s)}
\end{equation}
from which one deduces that the baryonic density is controled by the chemical
potential according to~: 
\begin{equation}
\rho=\int\,{4\,d^3 p\over (2\pi)^3} \,
\Theta\left(\mu\,-\,E^*_p\,-\,{g^2_\omega\over m_\omega^2}\,\rho\right)\label{RHOB}
\end{equation}
while the scalar density writes~:
\begin{equation}
\rho_S=\int\,{4\,d^3 p\over (2\pi)^3}\, {M^*_N\over E^*_p}\,
\Theta\left(\mu\,-\,E^*_p\,-\,{g^2_\omega \over m_\omega^2}\,\rho\right).
\end{equation}
The grand potential, which is obtained through a Legendre transform, can be 
written in the following form:
\begin{eqnarray}
\omega(\mu)&=&\varepsilon\,-\,\mu\,\rho\nonumber\\
&=&\int\,{4\,d^3 p\over (2\pi)^3} \left(E^*_p\,+\,{g^2_\omega \over
m_\omega^2}\,\rho\,-\,\mu\right)\,\Theta(\mu\,-\,E^*_p\,-\,{g^2_\omega \over
m_\omega^2}\,\rho)\,+\,V(s)\,-\,{g^2_\omega\over 2\, m_\omega^2}\,\rho^2.
\end{eqnarray}
Notice that the minimization (equation \ref{MIN}) can be equivalently
obtained from the condition $(\partial\omega/\partial\bar s)_\mu=0$. 
In order to derive the condensate and susceptibility we point out  that,  
in the context of this model, what plays the role of the chiral symmetry 
breaking parameter  is the quantity $c=f_\pi m_\pi^2$ which enters the symmetry breaking 
piece of the potential. Hence~:
\begin{equation}
\langle\bar q q\rangle={1\over 2} \left({\partial \omega\over\partial m}\right)_\mu=
{1\over 2} {\partial c\over\partial m}\left({\partial \omega\over\partial c}\right)_\mu
=-{1\over 2}{\partial c\over\partial m} \,\bar S\simeq {\langle\bar q q\rangle_{vac}\over
f_\pi}\,\bar S
\end{equation}
where we hase used the Feynman-Hellman theorem and the explicit expression of 
$\partial c/\partial m$ given by the model to leading order in the quark mass $m$.
Accordingly the in-medium scalar susceptibility is given by~:
\begin{equation}
\chi_S=\left({\partial\langle\bar q q\rangle\over\partial m}\right)_\mu=
-{1\over 2}\left({\partial c\over\partial m}\right)^2\,\left({\partial\bar S \over\partial c}\right)_\mu
\simeq -2\,{\langle\bar q q\rangle_{vac}^2\over f_\pi^2}\,\left({\partial\bar S \over\partial
c}\right)_\mu .\label{CHIS}
\end{equation}
The derivative $(\partial\bar S /\partial c)_\mu$ is 
obtained by taking the derivative of the
minimization equation (\ref{MIN}) with respect to the parameter $c$.
This gives~:
\begin{equation}
m^{*2}_\sigma \left({\partial\bar S \over\partial c}\right)_\mu=1\,-\,
g^*_S\, \Pi_0(0)\left[g^*_S \,{M^*_N\over E^*_F} \left({\partial\bar S \over\partial c}\right)_\mu
\,+\,{g^2_\omega\over  m_\omega^2}\left({\partial \rho \over\partial c}\right)_\mu\right] ,
\label{PASDENOM}\end{equation}
with $\Pi_0(0)= - 2 M^*_N\, p_F/\pi^2$. Notice that  $\Pi_0(0)$ is nothing but the 
non-relativistic free Fermi gas particle-hole polarization propagator 
in the Hartree scheme, at zero energy in the limit of vanishing momentum. The derivative of 
the baryonic  density is obtained by taking the derivative with respect to $c$ of 
eq. \ref{RHOB}, with the result~:
\begin{equation}
\left({\partial \rho \over\partial c}\right)_\mu=g^*_S
\left({\partial\bar S \over\partial c}\right)_\mu
\Pi_0(0)\left(1\,-\,{g^2_\omega\over  m_\omega^2}\, {E^*_F\over M^*_N}\,\Pi_0(0)\right)^{-1}.
\end{equation}
It follows that~:
\begin{equation}
\left({\partial\bar S \over\partial c}\right)_\mu=
{1\over  m^{*2}_\sigma}\,-\,{1\over  m^{*2}_\sigma}\, \Pi_{SS}(0)\,{1\over  m^{*2}_\sigma}
\label{CHISEFF}\end{equation}
where $\Pi_{SS}(0)$  is the full scalar polarization propagator 
(in which we include the coupling constant)~:
\begin{equation}
\Pi_{SS}(0)=g^{*2}_S \,{M^*_N\over E^*_F}\,\Pi_0(0)\,
\left[1 -\,\,\left({g^2_\omega\over  m_\omega^2}\, {E^*_F\over M^*_N}\,
- \,{g^{*2}_S\over m^{*2}_\sigma}\, {M^*_N\over E^*_F} \right) \Pi_0(0)\right]^{-1}.\label{PISS}
\end{equation}
We now comment these results. Firstly notice that the quantity 
$(\partial\bar S /\partial c)_\mu$, as written in eq. \ref{CHISEFF},  is  the in-medium 
scalar meson propagator (up to a minus sign) dressed by 
$NN^{-1}$ excitations. Moreover  it is satisfactory to realize that the expression of
$\Pi_{SS}$ (eq. \ref{PISS}) coincides with the one derived from the RPA equations 
in the ring approximation. In RPA the residual interaction  can be due either to the scalar 
meson exchange or to the vector one. In the latter case the mixed propagator $\Pi_{SV}$ 
enters.  The RPA equations read~:
\begin{eqnarray}
\Pi_{SS} &=& \Pi^0_{SS}\,+\,\Pi^0_{SS}\,D^0_S\,\Pi_{SS}\,-\,
\Pi^0_{SV}\,D^0_V\,\Pi_{VS}\nonumber\\
\Pi_{VS} &=& \Pi^0_{VS}\,+\,\Pi^0_{VS}\,D^0_S\,\Pi_{SS}\,-\,
\Pi^0_{VV}\,D^0_V\,\Pi_{VS}
\end{eqnarray} 
with $D^0_S=-1/m^{*2}_\sigma$ and $D^0_V=-1/m^{2}_\omega$. The solution for $\Pi_{SS}$ is~: 
\begin{equation}
\Pi_{SS}={\Pi^0_{SS}\,-\,D^0_V\,
\left(\Pi^0_{SV}\Pi^0_{VS}\,-\,\Pi^0_{SS}\,\Pi^0_{VV}\right)
\over 1\,-\,
\Pi^0_{SS}\,D^0_S\,+\,\Pi^0_{VV}\,D^0_V\,+\,
D^0_S\,D^0_V\,\left(\Pi^0_{SV}\Pi^0_{VS}\,-\,\Pi^0_{SS}\,\Pi^0_{VV}\right)} .
\end{equation}
One recovers our result of eq. \ref{PISS} since, as shown in the appendix,
 it is possible to check that:
\begin{equation}
\Pi^0_{SS}(0)=g^{*2}_S \,{M^*_N\over E^*_F}\,\Pi_0(0),
\quad\Pi^0_{VV}(0)=g^2_\omega\,{E^*_F\over M^*_N}\,\Pi_0(0),\quad
\Pi^0_{SV}(0)=\Pi^0_{VS}(0)=g^{*}_S \,g_\omega\,\Pi_0(0).\label{APP1}
\end{equation} 
In our approach the  Landau-Migdal parameter $F_0$ enters the RPA denominator of 
eq. \ref{PISS} which, at $\rho_0$, writes $1\,+\,F_0$.
This equation also shows that our residual particle-hole interaction is 
density dependent, in particular through the density dependence of 
$g^{*}_S$ and $m^*_\sigma$. Our approach provides 
a consistent relativistic frame in a chiral theory to derive this density dependence.
 A  similar expression has been given in ref. \cite{BT01}.
\newline
As for the full vector polarization propagator $\Pi_{VV}$, solution of the RPA equations,
it is given by~:
\begin{equation}
\Pi_{VV}(0)=g^{2}_\omega \,{E^*_F\over M^*_N }\,\Pi_0(0)
\left[1\,-\,\left({g^2_\omega\over  m_\omega^2} \,{E^*_F\over M^*_N}
- {g^{*2}_S\over m^{*2}_\sigma} \,{M^*_N\over E^*_F} \right) \Pi_0(0)\right]^{-1}.\label{PIVV}
\end{equation}
Notice that the same RPA denominator enters both expressions of $\Pi_{SS}$ and $\Pi_{VV}$, 
with omega and sigma exchange on the same footing.
The quantity  $\Pi_{VV}$, {\it i.e.}, the response to a probe which couples to the nucleon
density fluctuations,  is related to the nuclear compressibility. We introduce the 
incompressibility factor K of the nuclear medium, defined as~:
\begin{eqnarray}
K&=&9\,\rho\,{\partial^2\varepsilon\over\partial\rho^2}=9\,\rho\,{\partial\mu\over\partial\rho}
=9\,\rho\,{\partial\over\partial\rho}\left(E^*_F\, +\,{g^2_\omega\over m^2_\omega}\,\rho \right)
\nonumber\\
&=&{p_F\over E^*_F}\,{\partial p_F\over\partial\rho}\,+\,{M_N^*\over E^*_F}\, g^*_S\,
{\partial \bar s\over\partial\rho}\,+\,{g^2_\omega\over m^2_\omega} .
\end{eqnarray} 
The minimization (equation  \ref{MIN}) establishes the dependence of $\bar s$ with respect 
to $\rho$. By taking its derivative, one gets~: 
\begin{equation} 
{\partial \bar s\over\partial\rho}=-{ g^{*2}_S\over m^{*2}_\sigma}\,{M^*_N\over E^*_F} .
\end{equation}
It follows that~:
\begin{equation}
{\Pi_{VV}(0)\over g^2_\omega} = - {9\rho\over K} .
\end{equation}
This relation is well known  in the non relativistic situation. We have shown  that it 
 also applies in the relativistic case with $\Pi_{VV}$  given above (eq. \ref{PIVV}). 
At low densities 
where relativistic effects are small, there is no distinction between the scalar and 
vector propagators and we have $\Pi_{SS}(0)/g^{*2}_S\simeq\Pi_{VV}(0)/g^2_\omega$. 
More generally the relation is~: 
\begin{equation} 
{\Pi_{SS}(0)\over g^{*2}_S}=\left({M_N^*\over E^*_F}\right)^2\,{\Pi_{VV}(0)\over g^2_\omega} .
\end{equation}
With our values of parameters the parenthesis represents a reduction of about 10\% at $\rho_0$. 
The nuclear physics information on the nuclear
matter compressibility leads to a value of $K$ in the range $200-300$ MeV. This is close to the 
free Fermi gas value, which is  230 MeV for $M^*_N =M_N$. Thus the nuclear phenomenology
which constraints the residual interaction at saturation density, 
also constraints  scalar quantities, such as the scalar nuclear response and the QCD scalar
susceptibility. 
\section{Numerical results and discussion}
\subsection{Sigma mass}
The first consequence we wish to discuss is the  much debated problem of the density 
dependence of the scalar meson mass. Its explicit form  (in the chiral limit) is given by 
eq. \ref{MSIGMA} and we now come to the quantitative estimate. It is possible to get a 
first evaluation of the nucleon response, $\kappa_{NS}$, from the parameters of the 
QMC model, as given in \cite{GSRT96}. It gives for the dimensionless parameter 
$C =(f^2_\pi /2 M_N)\kappa_{NS}$  the value $C=0.45$. 
Two independant parameters remain to be fixed~: $g_S/m_\sigma$ and $g_\omega/m_\omega$.
Since $g_S=M_N/f_\pi$ is given  by our model and since  we take for the omega mass the vacuum
value $m_\omega=783$ MeV, the parameters to be fixed are in fact 
$m_\sigma$ and $g_\omega$. In the  first step of the discussion we keep for the scalar 
potential,  $V_0(s)$, only the quadratic part, namely,  $V_0(s)=m_\sigma^2 s^2/2$, 
as in the original Walecka 
or QMC model which both ignore the chiral softening of the sigma mass. In such a case 
the saturation properties ($\rho_0=0.17 fm^{-3}$, $E/A=-15.3$ MeV) are obtained by taking 
$m_\sigma=715$ MeV and  $g_\omega=7.47$. The corresponding incompressibility is 
$K=260$ MeV and the effective nucleon mass at $\rho_0$  is $M^*_N=760$ MeV. 
In general with the
introduction of $\kappa_{NS}$ the effective nucleon mass is closer to the free value 
than in pure Quantum Hadrodynamics where $M^*_N\simeq 600$ MeV.

If instead we  introduce the full chiral $V_0(s)$,  the effect of chiral restoration 
associated with the dropping of the sigma mass (see eq. \ref{MSIGMA} and the accompanying
discussion) tends to destroy the saturation mechanism. One possible way to recover
saturation is to increase the nucleonic susceptibility  $\kappa_{NS}$ (or equivalently 
the dimensionless parameter $C$). Indeed with a larger value, $C=0.8$, and keeping the other 
parameters to the same values we find saturation at $\rho_0$. However the binding energy 
$E/A=-17.8$ MeV and the incompressibility  $K=360$ MeV are too large. A  slight 
readjustment of the parameters $C=0.85$, $m_\sigma=750$ MeV  and $g_\omega=7$ can solve 
the problem for 
the binding energy (see dot-dashed curve of figure 1) but the incompressibility remains too 
large (above $300$ MeV). 
The fact that we do not fully reproduce with this simple  model the saturation properties 
is not surprising. We have assumed a linear response of the nucleon ($\kappa_{NS}$ is 
independent of $\bar s$), which is not founded. It is easy to introduce a field 
dependence. We take the simplest linear one, imposing the vanishing of the susceptibility 
at chiral restoration ($\bar s=-f_\pi$)~:
\begin{equation}
\kappa_{NS}\quad\to\quad\kappa_{NS}(\bar s)={\partial^2M_N\over\partial \bar s^2}=
\kappa_{NS}\left(1\,+\,{\bar s\over f_\pi}\right).
\end{equation}
Accordingly the effective nucleon mass becomes~:
\begin{equation}
M^*_N=M_N\,\left( 1+{\bar s\over f_{\pi }}\right)\,+{1\over 2}\,\kappa_{NS}\,
\bar s^2\left(1\,+\,{\bar s\over 3\,f_\pi}\right) .
\end{equation}
In this case the set of parameters~: $g_\omega=6.8$, $m_\sigma=750$ MeV and $C=1$, leads 
to correct saturation properties, with an incompressibility value $K=270$ MeV 
(full curve of figure 1) and $M^*_N(\rho_0)=760$ MeV. Notice that in this case the 
corresponding value of the residual interaction $F_0$ at $\rho_0$ is practically zero 
($F_0\simeq 0.03$). This is due to a delicate cancellation  between two large terms~: 
the omega exchange and the in-medium modified scalar exchange, the magnitude of each term 
being of the order of $3$.

Coming back  to the sigma mass evolution, the general behaviour is to 
a large extent independent of the exact field dependence of the susceptibility. 
Fig. 2 represents this evolution without (dot-dashed curve)  and with the effect of 
the nucleonic scalar response, introducing or not its field dependence. 
The two curves with this nucleonic 
scalar response are flat. The nucleon reaction largely  suppresses the strong softening 
due to chiral restoration which, if taken alone, would not be compatible with saturation
properties. A similar conclusion was reached in ref. \cite{BT01}. 

\begin{figure}                     
\centering
\includegraphics[width=0.5\linewidth,angle=270]{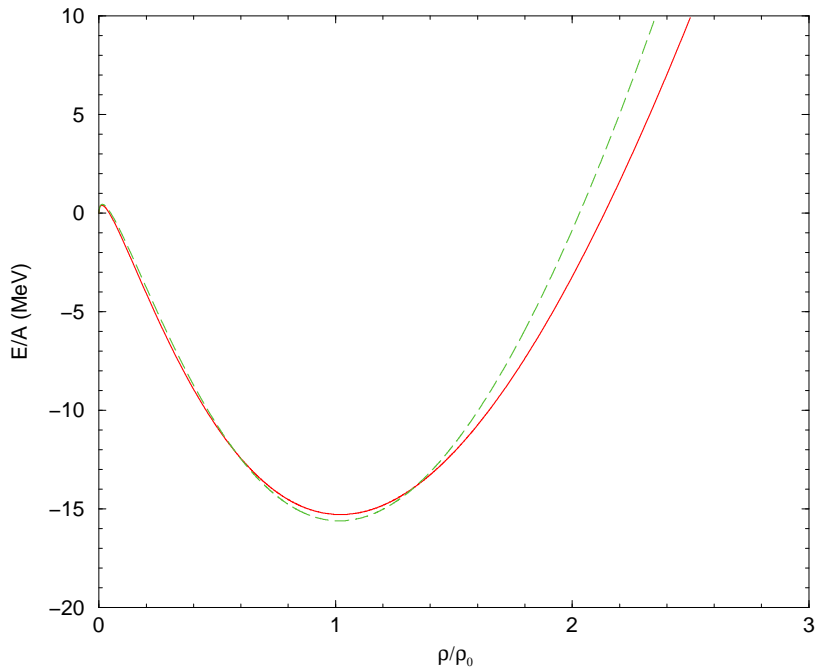} 
\caption{Binding energy of nuclear matter. The dashed line corresponds to the set of parameters
    $g_\omega=7$, $m_\sigma=750$ MeV and $C=0.85$ in the absence of the density
    dependence of the  nucleon susceptibility. The full
    line  corresponds to a  set of parameters 
  $g_\omega=6.8$, $m_\sigma=750$ MeV and $C=1$ with an explicit field (density) dependence
  of the nucleon susceptibility as explained in the text.}
\label{}
\end{figure} 
\begin{figure}                      
\centering
\includegraphics[width=0.5\linewidth,angle=270]{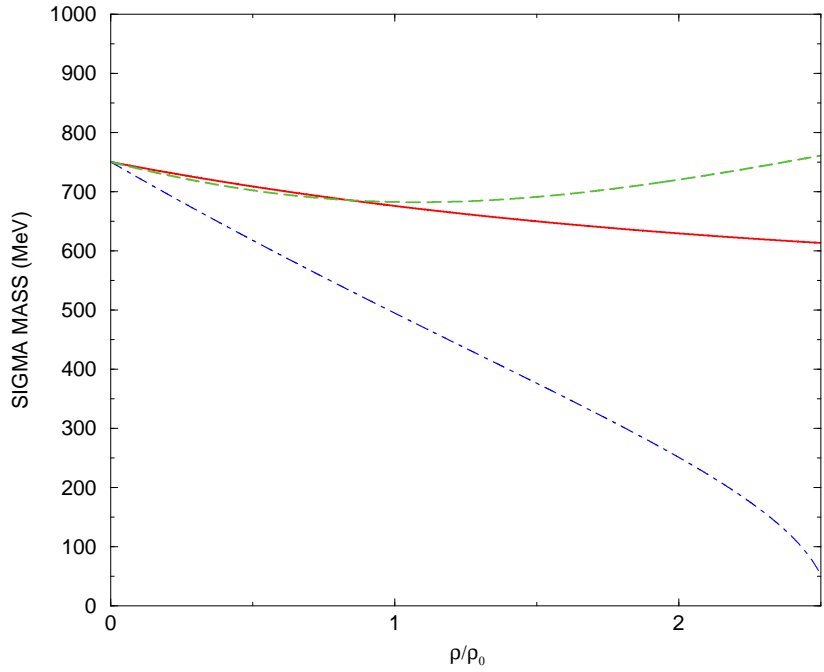} 
\caption{Density evolution of the sigma mass. Dotted line~: in 
  the absence of the field (density)
  dependence of  the nucleon susceptibility with  values of the parameters $g_\omega=7$, 
  $m_\sigma=750$ MeV and $C=0.85$. Full line~: with  density dependence of  the nucleon 
  susceptibility with  $g_\omega=6.8$, $m_\sigma=750$ MeV and 
  $C=1$. Dot-dashed line~:it corresponds to the case where only the chiral softening is
  included, without the effect of the nucleon susceptibility.}
\label{}
\end{figure} 
\begin{figure}                    
\centering
\includegraphics[width=0.5\linewidth,angle=270]{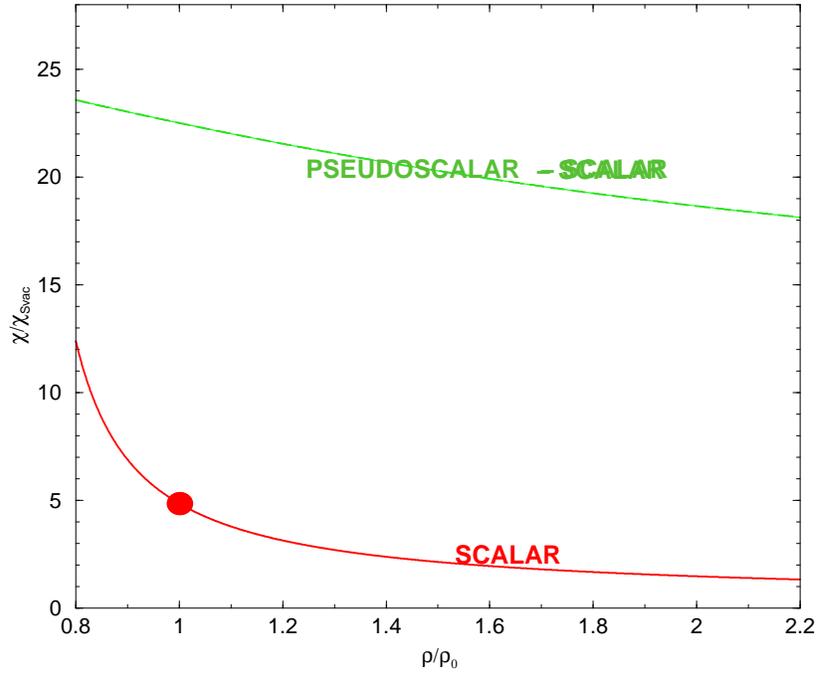} 
\caption{Density evolution of the QCD susceptibilities normalized to the
  vacuum value of the scalar one calculated with the field dependence of the nucleon
  susceptibility. Full curve~: scalar susceptibility. Dashed curve~: difference 
  between scalar and pseudoscalar susceptibilities.}
\label{}
\end{figure}

\subsection{Nuclear responses}
Our second point concerns the nuclear response to a probe which couples to the nucleon 
density, scalar or vector, for which we have given the relativistic expressions (eq.
\ref{PISS},\ref{PIVV}). 
At $\rho_0$ we have seen that the value of the incompressibility $K$ of nuclear matter  
lies in the range 200-300 MeV.  This is close to the free Fermi gas value, which is  
$K=230$ MeV for $M^*_N=M_N$. In QMC where the effective mass does not differ so much from the
free one, the experimental value of $K$  implies a small 
residual force at $\rho_0$. Our choice of parameters takes this constraint into account. 
This smallness of $F_0$ 
results from the accidental cancellation between omega and sigma exchanges. However this 
delicate balance is upset with a change of the density. Several factors are responsible 
for this phenomenon. Firstly the relativistic factor $(M^*_N/E^*_F)$ lowers the sigma 
exchange while it enhances the omega one, by a density dependent amount. In QHD, where the
effective nucleon mass is small, this influence is large. In QMC instead, with larger mass values
this is not the main effect. The main one is due to the action of the nucleonic 
reaction $\kappa_{NS}$ which is responsible for the decrease of the scalar coupling 
constant with increasing density. The sigma effectively decouples from the nucleon at 
large density, leaving the repulsive omega interaction to dominate. The reverse occurs 
at smaller density with the increase of $g_S$. The sigma attraction then fully develops 
and dominates the repulsive omega exchange. Thus, with increasing density, the residual 
interaction turns from attraction into repulsion. Each component being large the evolution 
is fast. For instance, with our parameters the resulting attraction is strong enough to 
produce a singularity of the polarization propagator at a density $\rho\simeq 0.6\rho_0$. 
We identify it with the spinodal instability. In the density region below the saturation 
one, the responses, scalar or vector, are collective with a resulting softening of the 
the response. It is reflected as en enhancement in magnitude of the nuclear susceptibility 
(proportional to $\Pi(0))$. Above $\rho_0$ instead collectivity hardens the response and 
decreases the susceptibility.
  
The density dependence of the residual interaction is well established 
\cite{ACPS83, NKS84}.
It is small in the nuclear interior and strongly attractive at the nuclear 
periphery. On the other hand the collective nature of the vector response is confirmed by  
data, from $(e,e')$ scattering, on the longitudinal response of various nuclei at small 
momenta. The charge response is a sum of a vector-isoscalar response and a vector-isovector 
one. Only the first part is relevant for our discussion. 
The data at small momenta (200-300 MeV/c) display a strong softening effect
with respect to the free response. Alberico {\it et al} \cite{ACEM87}
have attributed this feature to the collective caracter, at the nuclear periphery, of 
the isoscalar part of the charge response, in accordance with the present views.
The density dependence of the residual force with the non-trivial evolution from attraction 
into repulsion with increasing density naturally follows in our description with  sigma and 
omega exchange and with the incorporation of the nucleonic scalar response, 
necessary to favor the omega contribution relative to the sigma one with increasing density.

\subsection{QCD  susceptibilities} 
It is normal that the collective character of the nuclear response shows up  in  the 
QCD susceptibility, as the quark density fluctuations are coupled to the nucleonic ones. 
In ref. \cite{CE03}   we have derived, in the linear sigma model, the following relation 
between the in-medium QCD scalar susceptibility and the vacuum one~:
\begin{equation}
{\chi_S\over\chi_{S,vac}}={m^2_\sigma\over m^{*2}_\sigma}\left(1\,-\, 
{\Pi_{SS}(0)\over m^{*2}_\sigma}\right) .
\end{equation}
For a quantitative evaluation we calculate the r.h.s. in the model which incorporates 
the nucleonic response with its field dependence. The corresponding parameters are those
given previously, which reproduce the saturation properties. The resulting density
evolution is shown in fig.3. At $\rho_o$ the enhancement over the vacuum value is 
a factor of about 5. At lower density it becomes even larger due to 
the collectivity of the polarization propagator $\Pi_{SS}$, with the
attractive p-h force which enhances its magnitude. At larger densities instead the 
p-h force becomes repulsive which shows up in the gradual  disappearance of the medium effect, 
as depicted in fig. 3.
\newline
 It is interesting to contrast the evolution of the scalar 
susceptibility with that of the other QCD  susceptibility, the pseudoscalar one. This 
last quantity is linked to the fluctuations of 
the quark pseudoscalar-isovector density. The question is wether the increase in magnitude 
of the scalar susceptibility due to the coupling to the nucleon-hole 
states leads to a convergence effect between the two susceptibilities, which would be a signal
 for chiral symmetry restoration since the two susceptibilities become equal in the restored
 phase. In ref. \cite{CE03}   
we have demonstrated that the evolution of the pseudoscalar susceptibility follows that 
of the quark condensate, with~:
\begin{equation}
\chi_{PS}={\langle \bar q q (\rho)\rangle\over m_q}
\end{equation}
which diverges in the chiral limit, as it should.
The condensate on the r.h.s. is a function of the density and also of the quark mass. Since
 we are not
considering here pion loops which would produce a non-analytical behavior in $m_q$, far from
the restoration density we can expand the condensate, writing~:
\begin{equation}
\chi_{PS}={\langle \bar q q (\rho)\rangle\over m_q}=
{\langle \bar q q (\rho, m_q=0)\rangle\,+\,\chi_{S}\,m_q\over m_q} .
\end{equation}
This equation   provides in fact the difference between the pseudoscalar and 
the scalar susceptibilities. The important point is that this difference which
is governed by the  value of the condensate in the chiral limit 
is a smooth fuction of the density, which does not feel the large fluctuations near the 
spinodal instability. This means that the large increase of the scalar susceptibility 
 is also reflected in the pseudoscalar one, in such a way that the difference between the 
two remains smooth. We  have evaluated this difference within our model and it  is also
shown in fig. 3. It  starts to decrease linearly with 
the density but then flattens off, due to the action of the nucleonic reaction which 
limits the rise with density of the mean scalar field.  The convergence effect between 
the two susceptibilities is mild~:  with both quantities normalized to the vacuum scalar
one the difference between the two susceptibilities is $28$  in the vacuum and it becomes
$23$ at $\rho_0$. This evolution   reflects the restoration associated with the smooth 
drop of the condensate.  However the magnitude of the convergence effect is less marked that
previously obtained in our previous work \cite{CE03}.  
This difference with our previous
conclusion is due to the  interpretation of the condensate in
the expression of the pseudoscalar susceptibility. The scalar susceptibility term
of the quark condensate was not included in our previous work. 
\section{Conclusion}

In summary we have  formulated a relativistic effective theory of nuclear matter
based on the standard chiral symmetry properties. We have integrated in our approach
the concept of the nucleonic response to the nuclear  scalar field, which is necessary
to reach   a reasonable description of the saturation properties of nuclear matter, 
working at the Hartree level. Our present mean-field description  
ignores the role of the pion loop. For  a more complete description 
the corresponding correlation energy has to be included
and we expect some modification on our values of the parameters such as the bare sigma mass. 

In our framework we have also studied the relativistic RPA, scalar or vector isoscalar, 
responses of the nuclear medium. The residual interaction  is strongly density dependent. 
It turns from attraction into repulsion in the vicinity of the normal density, hence changing
the collective nature of the responses. 
The coupling of the quark density fluctuations to the nucleonic ones makes the QCD scalar
susceptibility sensitive to the collective behavior of the nuclear response. It 
is appreciably enhanced in the low density region where the residual ph force is attractive.
This collective behavior of the nucleon-hole polarization propagator
is also reflected in the pseudoscalar QCD susceptibility.

\appendix
\section{relativistic bare polarization propagators} 

First consider the bare first order polarization propagator in the vector channel~:
\begin{equation}
{\Pi^0_{VV}(0)\over g^2_\omega}=-\lim_{{\vec q}\to 0}
\int\,{8\,d^3 p\over (2\pi)^3} \,
{\Theta(p_F - p)\,\Theta(|\vec{p}+\vec{q}| - p_F )\over E^*_
{\vec{p}+\vec{q}}\,-\,E^*_p}.
\end{equation}
We multiply the denominator and the numerator by $E^*_ {\vec{p}+\vec{q}}\,+\,E^*_p$
whch leads to~:
\begin{equation}
{\Pi^0_{VV}(0)\over g^2_\omega}=-\lim_{{\vec q}\to 0}
\int\,{8\,d^3 p\over (2\pi)^3} \,(E^*_ {\vec{p}+\vec{q}}\,+\,E^*_p)\,
{\Theta(p_F - p)\,\Theta(|\vec{p}+\vec{q}| - p_F )\over (\vec{p}+\vec{q}^2)\,-\,p^2}.
\end{equation}
Obviously,  the factor $(E^*_ {\vec{p}+\vec{q}}\,+\,E^*_p)$ 
in the integrand can be replaced by $2\,E^*_F$ since in the limit ${\vec q}\to 0$,
 ${\vec p}$ has to lie on the Fermi surface.  It follows that~:
\begin{equation}
{\Pi^0_{VV}(0)\over g^2_\omega}=-{E^*_F\over M^*_N}\,\lim_{{\vec q}\to 0}
\int\,{8\,d^3 p\over (2\pi)^3} \,
{\Theta(p_F - p)\,\Theta(|\vec{p}+\vec{q}| - p_F )\over (\vec{p}+\vec{q}^2)/2M^*_N\,
-\,p^2/2M^*_N}
\end{equation}
and the remaining integral is manifestly the non relativistic $\Pi_0(0)$. This
establishes the second relation given in eq. \ref{APP1}. For each  scalar 
vertex one gets an additional multiplying factor $M^*_N/E^*_F$. Thus, for one scalar
vertex as is the case in the mixed term   $\Pi^0_{SV}(0)$, the relativistic
correction factor altogheter disappears, hence the relation  in eq. \ref{APP1}.
The presence of two scalar vertices leads to the first relation for the pure scalar
polarization propagator $\Pi^0_{SS}(0)$.

\section*{Acknowledgments} 
We thank P. Guichon and D. Davesne for fruitful discussions.



\begin{thebibliography}{99}
\bibitem{CEG02} G. Chanfray, M. Ericson, and P.A.M. Guichon, Phys. Rev C63 055202.
\bibitem{CE03} G. Chanfray, M. Ericson, EPJA (2003) 291.
\bibitem{SW86} B.D. Serot and J.D. Walecka, Adv. Nucl. Phys. 16 (1986) 1.
\bibitem{SW97} B.D. Serot and J.D. Walecka, Int. J; Mod. Phys. E16 (1997) 515.
\bibitem{BT01} W. Bentz and A.W. Thomas, Nucl. Phys. A696 (2001) 138. 
\bibitem{G88} P.A.M. Guichon, Phys. Lett. B200 (1988) 235.
\bibitem{GSRT96} P.A.M. Guichon, K. Saito, E. Rodionov and A.W. Thomas, Nucl. Phys. A601
(1996) 349.
\bibitem{HA99} T. Hatsuda, T. Kunihiro and H. Shimizu, Phys. Rev. Lett. 82 (1999) 2840; 
D. Jido, T. Hatsuda and T. Kunihiro, Phys. Rev. D63 (2001) 011901 .
\bibitem{B96} F. Bonnuti {\it et al.}, Nucl. Phys. A677 (2000) 213; Phys. Rev. Lett. 77 (1996)
603; Phys. Rev. C60 (1999) 018201. 
\bibitem{ST00} A. Starostin {\it et al.}, Phys. Rev. Lett. 85 (2000) 5539.
\bibitem{M02}J.G. Messchendorp {\it et al.}, Phys. Rev. Lett. 89 (2002) 222302.
\bibitem{CEG03} G. Chanfray, M. Ericson, and P.A.M. Guichon, C68 (2003)035209
\bibitem{C02} G. Chanfray, Plenary talk given at PANIC02, Osaka (Japan) 28 Sept-
4 Oct 2002; to be published in Nucl. Phys. A.
\bibitem{ACPS83}M.R. Anastasio,L.S. Celenza, W.S. Pong and C.M. Shakin Phys. Reports, 100 
(1983) 328.
\bibitem{NKS84} K. Nakayama, S. Krewald and J. Speth, Phys Lett 145B (1984) 310.
\bibitem{ACEM87} W.M.  Alberico, P. Czerski, M.Ericson  and A. Molinari, Nucl. Phys A462 (1987) 
269.
\end{thebibliography}
\end{document}